\documentclass[traditabstract]{aa}
\usepackage{graphicx}
\usepackage{natbib}
\usepackage{txfonts}
\usepackage{url}

\def\unter#1{\hbox to #1{\hrulefill}}

\def\ie{{\it i.e.,~}}
\def\chandra    {\emph{Chandra\,}}
\def\xmm        {\emph{XMM\,}}
\def\vla        {\emph{VLA\,}}

\def\gmrt {\emph{GMRT\,}}

\def \degpoint {$^{\circ}$}
\def \secpoint {$^{\prime \prime~}$}
\def \minpoint {$^{\prime~}$}
\def \ql {\textquotedblleft}
\def \qr {\textquotedblright}

\def\ie{i.e.\,}
\def\eg{e.g.\,}

\begin{document}
\title{Discovery of large--scale diffuse radio emission and of a new galaxy cluster in the surroundings of MACS\,J0520.7-1328}

   \author{
   	G. Macario\inst{1}
   	 \and
          H.~T. Intema\inst{2,3}
          \and
          C. Ferrari\inst{1}
          \and
          H. Bourdin\inst{4}
           \and
          S. Giacintucci\inst{5,6} 
          \and
          T. Venturi\inst{7}
          \and
          P. Mazzotta\inst{4}
          \and
          I. Bartalucci\inst{4}
          \and
           M. Johnston-Hollitt\inst{9}
           \and
           R. Cassano\inst{7}
	 \and
	 D. Dallacasa\inst{7,8}
          \and
	G. W. Pratt\inst{10}
          \and
          R. Kale\inst{7,8}
          \and
          S. Brown\inst{11}                    
           }

   \institute{
   Laboratoire Lagrange, UMR 7293, Universit\'e de Nice Sophia-Antipolis, CNRS, Observatoire de la C\^ote d'Azur, 06300 Nice, France
         \and
     Jansky Fellow of the National Radio Astronomy Observatory, 520 Edgemont Road, Charlottesville, VA 22903-2475, USA 
          \and
        National Radio Astronomy Observatory, P.O. Box O, 1003 Lopezville Road, Socorro, NM 87801-0387, USA 
	\and
      Dipartimento di Fisica, Universit\`a degli Studi di Roma \ql Tor Vergata\qr, via della Ricerca Scientifica 1, 00133 Roma, Italy
	\and
	Department of Astronomy, University of Maryland, College Park, MD 20742, USA
	\and
	Joint Space-Science Institute, University of Maryland, College Park, MD 20742-2421, USA
	\and 
	INAF-Istituto di Radioastronomia, via Gobetti 101, I-40129 Bologna, Italy
 	\and
  	Dipartimento di Fisica e Astronomia, via Ranzani 1, I-40127 Bologna, Italy 
 	 \and
	School of Chemical \& Physical Sciences, Victoria University of Wellington, P.O. Box 600, Wellington 6140, New Zealand 
	\and
 	Laboratoire AIM, IRFU/Service d'Astrophysique � CEA/DSM � CNRS � Universit\'e Paris Diderot, B�t. 709, CEA-Saclay, 91191 Gif-sur-Yvette Cedex
	\and
	Department of Physics and Astronomy, University of Iowa, 203 Van Allen Hall, Iowa City, IA 52242, USA		 
	}

   \date{accepted for publication in A\&A }

  \abstract
   {We report the discovery of large--scale diffuse radio emission South-East of the galaxy cluster MACS\,J0520.7-1328, 
   detected through high sensitivity Giant Metrewave Radio Telescope 323~MHz observations.  
   This emission is dominated by an elongated diffuse radio source and surrounded by other features of lower surface brightness. 
    Patches of these faint sources are marginally detected in a 1.4 GHz image obtained through a re-analysis of archival NVSS data. 
   Interestingly, the elongated radio source coincides with a previously unclassified extended X-ray source. 
We perform a multi--wavelength analysis based on archival infrared, optical and X-ray \chandra data. 
We find that this source is a 
low--temperature ($\sim$3.6 keV) cluster of galaxies, with indications of a disturbed dynamical state,  
located at a redshift that is consistent with the one of the main galaxy cluster MACS\,J0520.7-132 (z=0.336). 
We suggest that  
the diffuse radio emission is associated with the non-thermal components in the intracluster and intergalactic medium 
in and around the newly detected cluster. We are planning deeper multi--wavelength and multi-frequency radio observations to
accurately investigate the dynamical scenario of the two clusters and to address more precisely the nature of the complex radio emission. 
} 
  
\authorrunning{Macario et al.}
\titlerunning{Discovery of diffuse radio emission and of a new galaxy cluster in the surroundings of MACS\,J0520.7-1328}
   \maketitle
%

\section{Introduction}


\begin{figure*}
   \centering
   \includegraphics[width=0.95\textwidth]{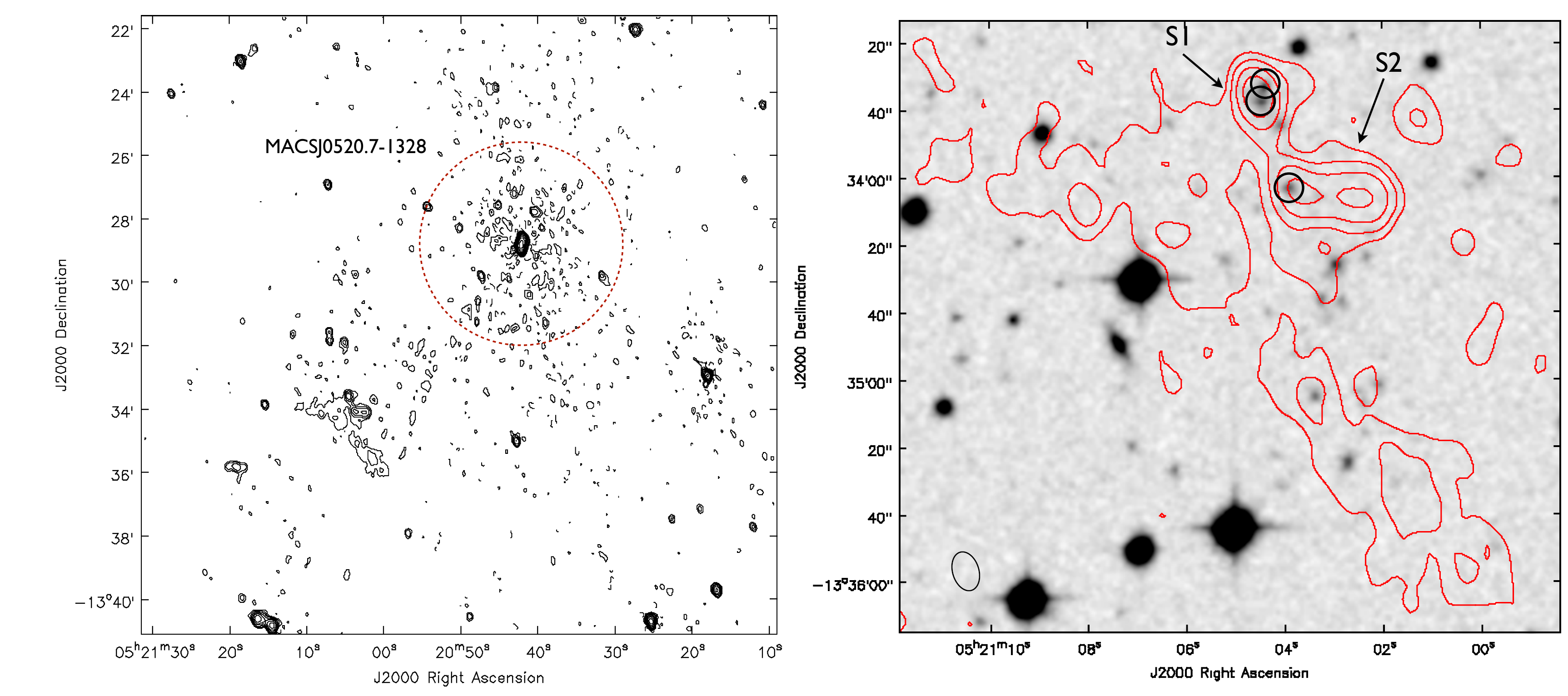}   
   \caption{
{\em Left panel:} 323~MHz GMRT full--resolution (11.8\secpoint $\times$ 7.8\secpoint, p.a. 16.9\degpoint) image of the 19\minpoint$\times$19\minpoint area around the cluster M0520. Contours are spaced by a factor of two and start at $\pm$3$\sigma$=0.33 mJy/b (negative contours are dashed). The  dashed circle indicates the central region of 1 Mpc radius around the cluster M\,0520 . {\em Right panel:} zoom in of the area around the diffuse elongated source; radio contours (same as  in {\em Left panel}) are overlaid on the DSS2 optical red image. 
The black circles show position of the three galaxies from DSS catalog, which are likely optical counterparts of radio sources S1 and S2.
}
         \label{fig:fig1}
   \end{figure*}



\begin{figure*}
   \centering
   \includegraphics[width=1.\textwidth]{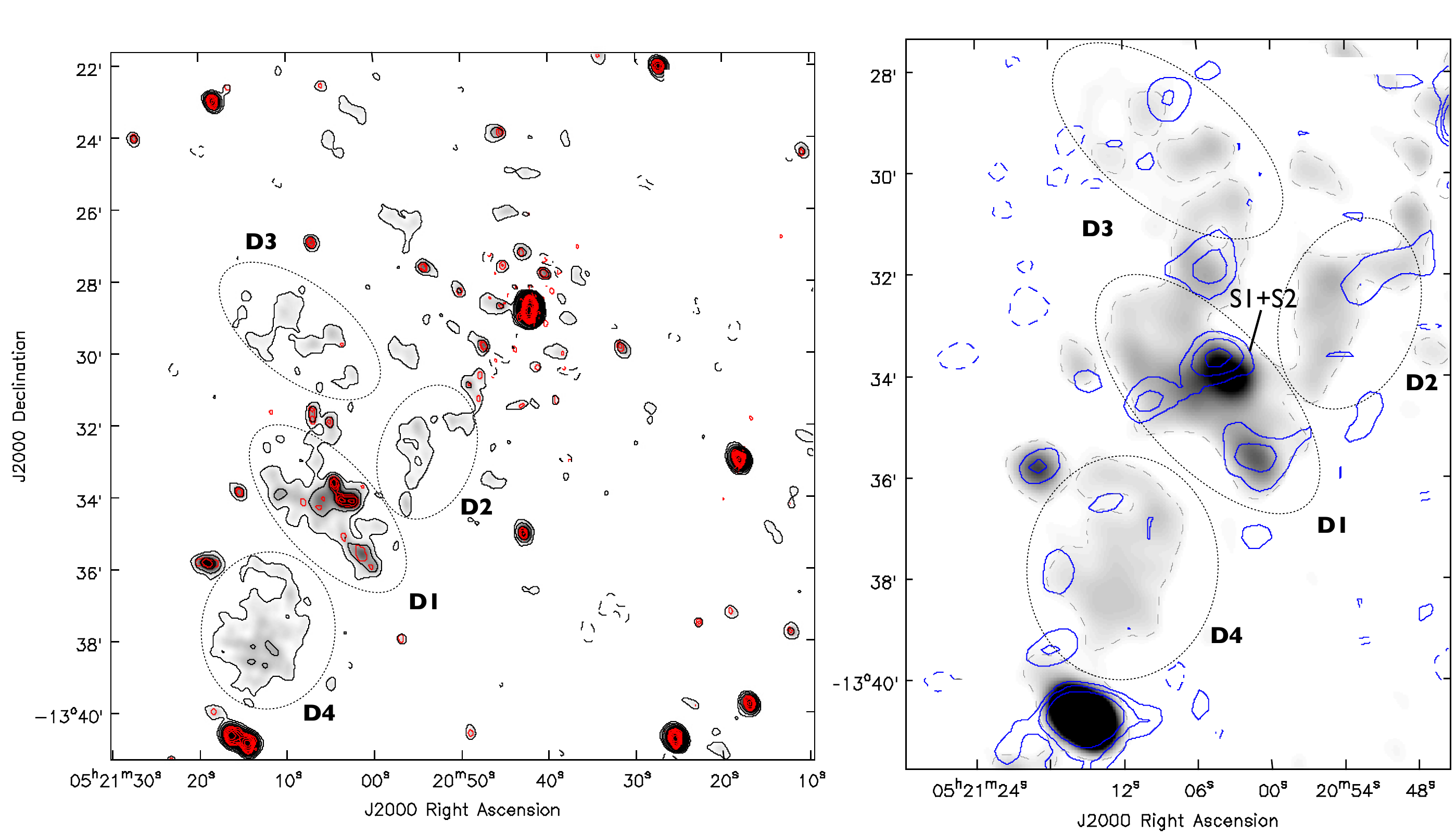}      
   \caption
   {
   {\em Left panel}: 323~MHz GMRT low--resolution (21.1\secpoint $\times$ 17.2\secpoint, p.a. 28.8\degpoint) image of the same 19\minpoint$\times$19\minpoint region around the cluster MACS\,J0520 (same as in Fig. \ref{fig:fig1}); black contours are spaced by a factor of two and start at 
   $\pm$3$\sigma_{323\rm MHz}$=0.82 mJy/b (negative contours are dashed). To highlight the position of discrete compact radio sources in the field, 
   contours from the full--resolution image (see Fig. \ref{fig:fig1}) are over--plotted in red, starting from $\pm6\sigma$=0.66 mJy/b. 
   Dotted ellipses mark the location of diffuse emission, labelled D1 to D4. 
    {\em Right panel:}  \vla 1.4 GHz contours (in blue) of an image obtained from re--processed NVSS data (45.8\secpoint $\times$ 29.9\secpoint, p.a. -70.0\degpoint); 
    contours start at $\pm2\sigma_{1400\rm MHz}=0.5$ mJy/b and are spaced by a factor of 2  (blue dashed are negative). The contours are overlaid on a 323~MHz GMRT image 
    of similar resolution (41.5\secpoint$\times$33.5\secpoint, p.a. 21.3\degpoint), shown in grey scale (the grey dashed contour corresponds to 3$\sigma_{323\rm MHz}$=1.65  mJy/b).
    }
         \label{fig:fig_2}
   \end{figure*}
   

In the currently supported hierarchical scenario of structure formation,  
galaxy clusters assemble and evolve through mergers and accretion of smaller units of matter, 
at the intersection of cosmic filaments connecting clusters into the  large--scale structure of the Universe. 

In addition to the most common methods for clusters detections 
-- based on the study of 
galaxy over-densities, lensing, extended X-ray emission due to thermal bremmstrahlung from the intracluster medium (ICM), 
and the sub-mm signal due to Inverse Compton scattering of CMB photons by hot intracluster electrons  (SZ effect) -- 
observations in the radio band can also probe the existence of clusters and/or large--scale structures. 
Powerful high-redshift radio galaxies are known to be efficient tools for detecting galaxy clusters at high redshifts (z$>$1.5; 
\eg \citealt{galametz13}, and reference therein). 
At lower redshift, tailed radio galaxies are being extensively and successfully used for cluster searches.  
Since they are known to be associated with galaxy clusters, they can be used to trace the high density environments in the Universe (\eg \citealt{gv09}).  
Several previously unobserved clusters have been actually identified thanks to the detection of tailed sources
(\eg  \citealt{blanton03}; \citealt{smolcic07}; \citealt{gv09}; \citealt{kantharia09}; \citealt{mao2010}). \\
Beyond tailed radio galaxies, it is nowadays known that a fraction of merging clusters 
host diffuse Mpc--scale radio emission (such as radio halos and relics, that are named according with their location and observational properties; \eg  see \citealt{ferrari08}). 
This synchrotron emission is unrelated to individual cluster radio galaxies and directly probes the existence of 
non-thermal components (relativistic particles and magnetic fields) mixed with the thermal ICM.  
To date, all these sources are found to be associated with dynamically disturbed / merging systems (\eg \citealt{cassano10a}; \citealt{cassano13}; \citealt{feretti12}, for a recent review). 
The detection of such sources can become -- in principle -- a powerful tool to trace previously unidentified clusters. \\
Upcoming deep and wide-field radio surveys with new radio interferometers like LOFAR  (Low Frequency Array; \eg \citealt{vanhaarlem13}), 
ASKAP  (Australian Square Kilometer Array Pathfinder; \eg \citealt{johnston08}), MeerKAT (\eg \citealt{booth12}) and MWA   (Murchison Widefield Array; \eg \citealt{tingay2013}) will allow systematic searches for new galaxy clusters based on the detection of tailed radio galaxies and diffuse extended emission 
(see \eg \citealt{ensslin02}; \citealt{cassano10b}). 
For these studies, LOFAR and MWA are particularly suitable, thanks to the very large field of view covered at low radio frequencies that makes them a powerful survey telescope (see \ie the ongoing all-sky surveys LOFAR MSSS, \citealt{heald13}, and MWA GLEAM, \citealt{lss}, in preparation).\\
In this paper we present an example of a new cluster detection made possible thanks to deep low--frequency radio observations. 
We report the serendipitous discovery with the Giant Metrewave Radio Telescope (\gmrt) at 323 MHz of new large--scale diffuse radio emission in the proximity of the cluster MACS\,J0520.7-1328 (M\,0520 hereinafter). \\
M\,0520 is one of the X-ray brightest clusters of galaxies from the MAssive Cluster Survey (MACS, \citealt{Ebeling10}). It is located at $z$=0.336 and has an X-ray luminosity of L$_{\rm r500, [0.1-2.4 keV]} = 7.8 \times 10^{44}$ erg/s \citep{Mantz10}\footnote{Corrected for our slightly different choice of cosmological parameters.}. 
M0520 is classified as a relatively relaxed system on the basis of its morphological properties, \ie a good alignment between 
the galaxy and ICM distributions, and quite concentric X-ray surface brightness contours in its central 1.5 Mpc region \citep[morphology code 2 in][]{Ebeling10}. 
This cluster is also found to be a lensing system, though no mass estimate has been derived from lensing studies \citep{Horesh10}. 
M\,0520 is identified in the Planck SZ cluster catalogue as PSZ1 G215.29-26.09,  with a mass estimated from the SZ of M$_{500}\sim$6.15$\times$10$^{14}$ M$_{\odot}$ \citep{plk2013}. \\
For our analysis, we made use of our new \gmrt\ radio observations and of multi--wavelength data available from archives and the literature. 
The paper is presented as follows:  in Sect. 2 we describe the new \gmrt observations and data reduction, along with the analysis of the detected emission; 
in Sect. 3, archival \chandra X-ray data analysis of the cluster is presented; in Sect. 4, we report 
optical/IR analysis based on available public catalogues. A multi--wavelength comparison is reported in Sect. 5, and the results are discussed in Sect. 6. \\
The adopted cosmology is  $\Lambda$CDM, with ${\rm H}_0$=71 km ${\rm s}^{-1} {\rm Mpc}^{-1}$, $\Omega_{\rm m} = 0.27$, $\Omega_{\Lambda} = 0.73$. At the redshift of the cluster 1$^{\prime}$ corresponds to $\sim$290 kpc. 

\section{Radio analysis}
\label{sec:radio}

\subsection{Observations and data reduction}

M0520 was observed with the \gmrt at 323 MHz in November 2011, as part of an ongoing large observational programme ($\sim$350 hours of total observing time) named the MACS-Planck Radio Halo cluster Project. The project is devoted to the search for diffuse radio emission in a sample of galaxy clusters selected from the MACS (X--ray) and the \xmm-Planck (X-ray/SZ) catalogs in the  redshift range 0.3-0.45, which is relatively  unexplored in terms of deep radio observations. For  this project, we recently completed a deep pointed survey carried out mainly at low frequency (25 targets with the \gmrt at 323 MHz, 7 with the Australia Telescope Compact Array -- ATCA at 2.1 GHz), optimised for the detection of diffuse faint cluster radio emission. Results will be presented in a forthcoming paper (Macario et al. in preparation). \\ 

Data were recorded every 16.1~seconds,  with 256~frequency channels covering 32~MHz of bandwidth.  Data reduction was performed using the NRAO Astronomical Image Processing System\footnote{\url{http://www.aips.nrao.edu}} (AIPS) package and  Source Peeling and Atmospheric Modelling (SPAM) software \citep{Intema09}.  After data editing, the remaining effective bandwidth is 31.2 ~MHz, centred at 323~MHz.  The total effective time on source is  $\sim$5 hours. The flux scale and the bandpass shape were determined from $\sim$ 25 minutes observation of 3C\,147 at the end of observations, adopting a flux density of 52.2 Jy at 323 MHz (following the Perley-Taylor 1999 flux scale\footnote{\url{http://www.vla.nrao.edu/astro/calib/manual/}}). 
The amplitude calibration was applied to the target field data, followed by additional RFI flagging and frequency averaging to 24~channels of 1.3~MHz each. We phase-calibrated the target field with a simple point source model derived from the NVSS \citep{Condon98} and WENSS \citep{Rengelink97}, followed by several rounds of wide-field imaging, CLEAN deconvolution and self-calibration. 
 Only the final self-calibration round included amplitude calibration on a 1-minute time scale to correct for antenna gain amplitude variations. Before applying, the gains per antenna were normalized using only the target field scan closest in time to the flux calibrator scan, thereby enforcing the correct flux scale across the whole observation run. 
SPAM ionospheric calibration and imaging was applied to the data. The average residual amplitude errors are estimated to be $\lesssim$ 8\% (\eg, \citealt{Chandra04}). 
For more details on the data reduction procedure, we refer to \cite{Macario13}. \\
Once the calibration converged, a set of final images were obtained, starting from the full resolution and than progressively applying various parameters on data weighting to enhance the information on the diffuse emission, sampled by the shortest baselines.  

\subsection{Detection of diffuse extended radio emission}
\label{sec:detec_r}

Fig.\,\ref{fig:fig1} ({\em Left panel}) displays contours of the final full--resolution image at 323~MHz (obtained with uniform weighting of the visibilities), corrected for the primary beam. The resolution is 11.8\secpoint $\times$ 7.8\secpoint and the average rms noise level is  $\sigma_{\rm 323~MHz}$ = 0.11 mJy beam$^{-1}$. 
A 19\minpoint $\times$ 19\minpoint field is shown, centred about 3\minpoint South-East (SE) of the telescope pointing direction ({\it i.e.} the centre of the main cluster).  The radio map of  the central field of M\,0520 is totally dominated by a very bright and slightly extended radio source with a flux density of $\sim$0.5 Jy, associated with the brightest cluster galaxy (hereafter BCG). Residual emission around it is 
likely due to deconvolution artefacts originated by the relatively strong emission from the BCG, which limits the dynamic range over the central Mpc of the cluster area (dashed circle).

Very interestingly, a diffuse and elongated low surface brightness radio source is located about 8\minpoint SE of the cluster centre. 
It shows a quite  regular, nearly linear shape, and it has an extension over $\sim$2.8\minpoint along its major axis. 
We note that this emission is clearly not due to the artefacts caused by the strong central source, 
that mostly affect a limited area (radius $\sim$1.5\minpoint ) around the BCG. 
A magnification of the diffuse source is shown in the {\em Right panel} of Fig.\,\ref{fig:fig1}, 
with the radio contours overlaid on the DSS2 optical red image
\footnote{The M\,0520 field is not covered by the SDSS.}. 
Two radio sources, named S1 and S2, can be distinguished, and are likely associated with optical counterparts.  
Their flux densities are 6.4$\pm$0.5 mJy (for the point-like source S1) and 12.0$\pm$1.0 mJy (for the double source S2). 
The diffuse emission, on the other hand, lacks obvious optical counterparts. 
We note that a bridge of radio emission seems to connect source S1 with the northern part of the diffuse source.

As Fig. \ref{fig:fig1} ({\em Left panel}) shows, 
peaks of residual emission are spread North-West and South-East of this elongated source, 
suggesting the presence of additional underlying extended emission of lower surface brightness.  
To highlight this feature, as well as to better image the diffuse elongated source, 
we produced images at lower resolution, by using different weighting schemes and by applying Gaussian 
tapers to give more weight to the short baselines. \\
In the left panel of Fig.  \ref{fig:fig_2} we show a low--resolution image (grey scale and black contours) 
of the central area around M0520 (same as in Fig. \ref{fig:fig1}, {\em Left panel}), 
obtained with robust 0.5 weighting and a 10 k$\lambda$ Gaussian taper to down--weight long baselines. 
At this resolution (21.1\secpoint$\times$17.2\secpoint), the diffuse elongated source (labelled here as D1) 
appears slightly more extended, with a more homogeneous brightness distribution. 
This image also brings forth other patches of diffuse emission surrounding D1, 
with lower surface brightness and extended on large--scales. 
In particular, North-West of D1, a filamentary structure (labelled as D2) extends 
over $\sim$1.8\minpoint from the South-West part of D1 towards the centre of M0520.    
North of D1 other patches of emission (labelled as D3) are found, with similar orientation as D1 and extended over $\sim2.7$\minpoint. 
A more regular roundish patch ($\sim 2$\minpoint$\times$2\minpoint) 
of low surface brightness emission (labelled as D4) is located South-East of D1. \\
To check the reliability of this underlying emission (D2,D3,D4), 
we subtracted from the uv--data the contribution from all the unrelated radio sources, 
including  those from D1, and carefully inspected the residual dataset, 
especially at the short  baselines. Neither residual low level RFI, nor bad data 
were found. We imaged the residual dataset with different parameters, and found 
the emission is seen by all baselines shorter than 0.7k$\lambda$.  
We also produced images by selecting different groups of  
channels and time ranges, and found the emission is visible at all frequencies 
and for the whole duration observation.  
Such emission regions have only positive brightness values, and we can exclude that they are 
artefacts arising from the bright radio galaxy at the centre of M\,0520  (that affect a smaller area around it; \ie within the dashed circle in Fig. \ref{fig:fig1}). 
In particular, the extended emission regions D2 and D3 do not have (positive or negative) counterparts at the opposite side of the bright central source, 
which strongly argues against a connection with its residual sidelobes. \\
In Fig. \ref{fig:fig_2}, the red contours represent the full-resolution image 
(shown  in Fig. \ref{fig:fig1}, {\em Left panel}), starting at the level of $\pm6\sigma$, 
to help identifying the position of individual discrete radio sources. 
We note that no discrete radio sources are embedded in the patches of extended emission (D2, D3, D4), excluding 
any possibility that these features result from the blending of compact sources.

%
\begin{table*}[ht!]
\caption{Properties of the diffuse radio emission}   
\begin{center}
\begin{tabular}{lccccc}
\hline\hline
\multicolumn{1}{l}{}&\multicolumn{2}{c}{Integrated flux densities [mJy] }  & \multicolumn{1}{c}{Angular extent} & \multicolumn{1}{c}{Linear extent$^{*}$} & \multicolumn{1}{c}{$P_{323 \rm MHz}$$^{*}$}  \\
\multicolumn{1}{l}{}&\multicolumn{2}{c}{(323 MHz)\,\,\, (1400 MHz)} &   \multicolumn{1}{c}{( \minpoint $\times$ \minpoint)} & \multicolumn{1}{c}{(kpc$\times$kpc)} & \multicolumn{1}{c}{(W Hz$^{-1}$)}  \\
\hline
\multicolumn{1}{l}{D1}&\multicolumn{2}{c}{$61.2\pm5.7$ \,\,\,\,\,\,\,  $9.0\pm1.6$} &   \multicolumn{1}{c}{2.8$\times$1.4} &    \multicolumn{1}{c}{800$\times$400} & \multicolumn{1}{c}{$2.3\times10^{25}$}  \\
\multicolumn{1}{l}{D2}&\multicolumn{2}{c}{$32.0\pm3.6$ \,\,\,\,\,\,\, $7.2\pm1.6$} &   \multicolumn{1}{c}{1.8$\times$1.3}  & \multicolumn{1}{c}{520$\times$370} & \multicolumn{1}{c}{$1.2\times10^{25}$} \\
\multicolumn{1}{l}{D3}&\multicolumn{2}{c}{$34.7\pm4.0$ \,\,\,\,\,\,\, $1.9\pm1.4$} &    \multicolumn{1}{c}{2.7$\times$1.3}  & \multicolumn{1}{c}{780$\times$375} & \multicolumn{1}{c}{$1.3\times10^{25}$}  \\
\multicolumn{1}{l}{D4}&\multicolumn{2}{c}{$48.5\pm5.0$ \,\,\,\,\,\,\, $8.4\pm1.9$} &   \multicolumn{1}{c}{2.2$\times$1.9}  & \multicolumn{1}{c}{635$\times$550} & \multicolumn{1}{c}{$1.8\times10^{25}$}  \\
\hline
\end{tabular}
\end{center} 
$^{*}$ Assuming the emission is located at the redshift of M0520
\label{tab:tab_flux}
\end{table*}



\subsection{Analysis}
\label{sec:analy_r}

In search for confirmation of the newly detected diffuse emission, 
as well as to provide a rough estimates of its spectral index, we have inspected 
all the available images of the M0520 field in the public radio surveys. 
Due to their low sensitivities, none of the diffuse components (D1, D2, D3, D4; see Fig. \ref{fig:fig_2}, left) are either detected in the 74 MHz VLSS \citep{Kassim03,Lane08} and VLSSr \citep{Lane12}, nor in the 153 MHz TGSS\footnote{\url{http://tgss.ncra.tifr.res.in/150MHz/tgss.html}} surveys. \\
Hints of diffuse emission are visible in the NVSS (NRAO VLA Sky Survey; \citealt{Condon98}) 1.4 GHz image. 
In order to improve the quality of the survey image, 
we have reprocessed and analysed the NVSS (project AC308) pointing containing the M0520 field. 
The data were imaged after a new calibration and phase self-calibration in AIPS. 
The rms sensitivity level achieved in the final image is $\sim$0.25 mJy/beam, with a restoring beam of 
45.8\secpoint $\times$ 29.9\secpoint\,(p.a.  -70.0\degpoint), that is about 3 
times lower  than the 
NVSS public image (local noise of $\sim$0.5 mJy/beam, 45\secpoint\ restoring beam). 
The residual amplitude errors are $\sim$5\%.
In {\em Right panel} of Fig. \ref{fig:fig_2} we show the contours (in blue) from the final re-calibrated NVSS image 
(starting from $\pm2\sigma_{1400 \rm MHz}$=0.5 mJy/b level). 
For a proper comparison, these are overlaid on a 323 MHz \gmrt  image (in grey scale) of similar resolution 
(41.5\secpoint$\times$33.5\secpoint), obtained using only the visibility data corresponding to 
baselines shorter than 15k$\lambda$, 
with robust 0 weighting and a 4 k$\lambda$ Gaussian taper. 
The dashed grey contour corresponds to 
the $+$3$\sigma_{323 \rm MHz}$=1.65 mJy/beam level of the image. 
At the $\sim$2$\sigma_{1400 \rm MHz}$ level, 
the diffuse components clearly detected at 323 MHz are only marginally detected in the NVSS data.   
In particular, when considering the diffuse elongated source D1, 
only part of the emission South-East of S1 and its South-West brightest region are detected.  
Only patches of the emission within the regions D2, D3 and D4 are detected. 
This is due to the worse surface brightness sensitivity of the NVSS image compared to the 323 MHz image.  
We measured the total flux densities at 323 MHz and 1.4 GHz by 
integrating the two images, corrected for the respective primary beam, over the same areas defined by the dashed ellipses (see Fig. \ref{fig:fig_2}, {\em Right panel}). 
S1 and S2 are blended at the low--resolution of the 1.4 GHz image, not allowing an accurate measurement of their individual flux densities. 
We thus estimated 
their global flux density by integrating the two low--resolution images in Fig. \ref{fig:fig_2} ({\em Right panel}) 
over a same area that encompasses their emission at both frequencies; we found $\sim 4.5$ mJy and  $\sim 30$ mJy 
at 1.4 GHz and 323 MHz respectively. These values were subtracted from the total flux densities measured in the D1 region, 
to provide an estimate of the flux density of the diffuse emission only. 
\\
The flux density measurements of each component are reported in Table \ref{tab:tab_flux}, 
together with their uncertainties. These are computed as: \\
$\sigma_S=\sqrt{(\sigma_{rms}\times\sqrt{N_{beam}})^2 + (\sigma_{amp})^2}$ \\
where 
$\sigma_{rms}$ is the rms noise of the image, $N_{beam}$ 
is the number of independent beams in the region where the flux densities are measured, and $\sigma_{amp}$ is the 
residual amplitude error. \\
Due to the marginal detection in the NVSS, the measurements of the flux densities of each patch of 
emission at 1.4 GHz are actually affected by large uncertainties. 
Moreover, part of the flux density is likely missed in the reprocessed 
NVSS image, due to the very short duration ($\sim$1 minute) of the observation, 
that results in a sparse uv--coverage and loss of sensitivity to the structure. 
Therefore it is not possible to properly estimate the spectral indices for each component. 
We are planning deeper and higher resolution follow-up observations at 1.4 GHz and 
other frequencies, that are necessary for a precise spectral analysis 
and to perform an accurate study of the newly discovered extended emission.  

The angular extent of each component  (D1-D4), estimated measuring the major and minor axis of each ellipse of Fig. \ref{fig:fig_2}, 
are reported in Table \ref{tab:tab_flux}.  Under the assumption that those 
structures are located at the cluster redshift $z$=0.336, 
their  projected linear extents are of the order of 500-800 kpc (see Table \ref{tab:tab_flux}). 
The derived total radio powers at 323 MHz is also reported in Table \ref{tab:tab_flux}, under the same assumption.


\begin{figure}
   \centering
   \includegraphics[width=0.47\textwidth]{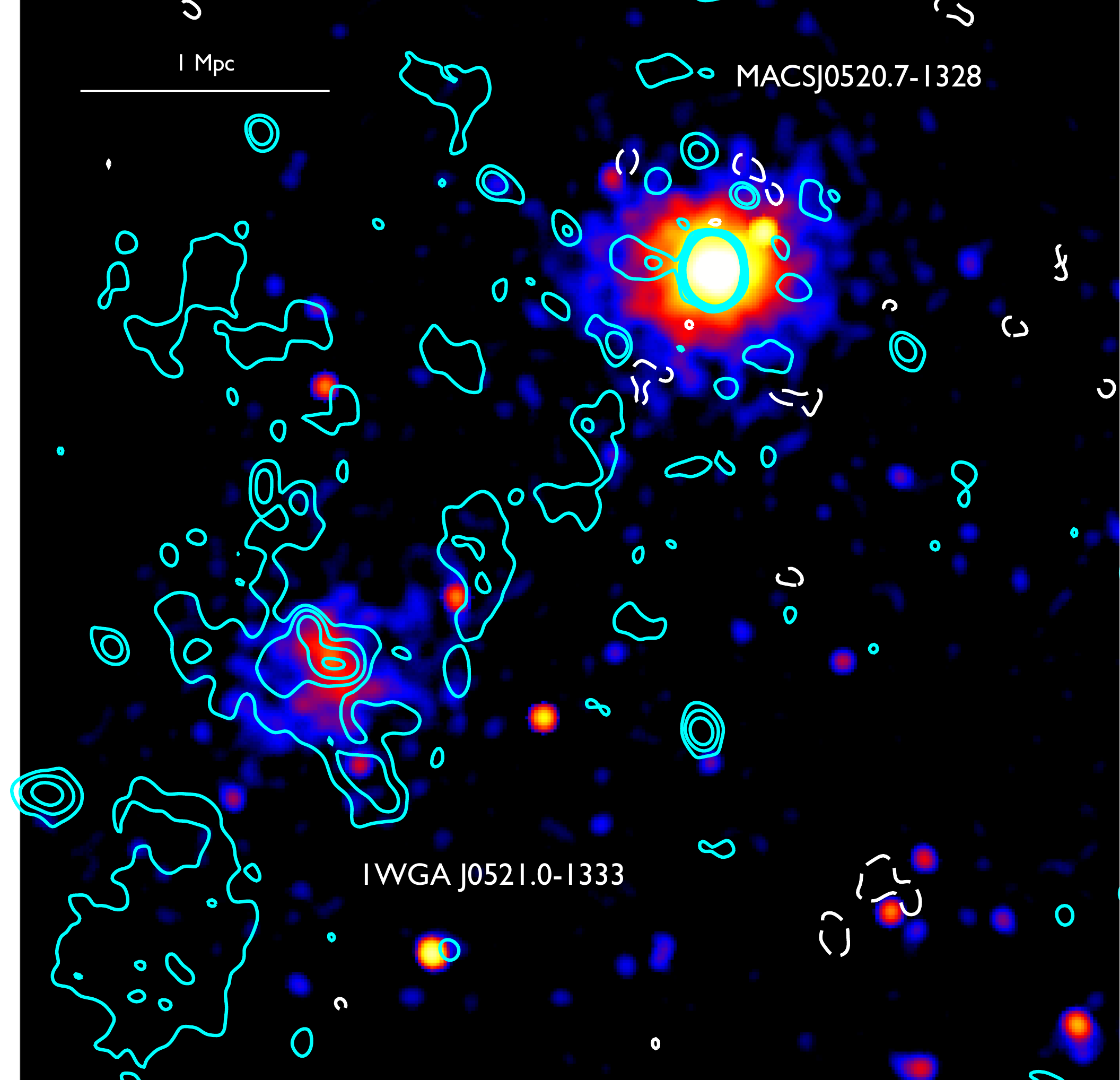}
      \caption{Smoothed \chandra X-ray image in the 0.5-2.5 keV energy band, with overlaid the 323 MHz \gmrt low--resolution contours 
      (same as in Fig. \ref{fig:fig_2}, {\em Left panel}; white are negative, cyan are positive). 
}
         \label{fig:fig3}
   \end{figure}

\section{\chandra X-ray analysis}

A simple overlay of the \gmrt radio contours on the archival \chandra\ raw image of M0520 central field ($\sim$ 20\minpoint $\times$ 20\minpoint) immediately revealed that the newly detected diffuse elongated radio source D1 (see Sect. \ref{sec:analy_r})  coincides with an extended X-ray component, unclassified in the literature. 
The nature of this diffuse X-ray emission is studied in the following sections through available multi-wavelength data.

\subsection{Data reduction and images}
\label{sec:Xdatar}

We have re--analysed the \chandra ACIS-I X-ray data from the public archive 
(obsid: 3272, total exposure $\sim$20 ks, same as the one analysed in \citealt{Mantz10}). 
A filtering of the hard and soft event light curves only
reduced this exposure by a 1\% factor. We have binned all photon events in
sky coordinates and energy with a fixed angular resolution of 2.5\secpoint, 
and a variable energy resolution matching the camera
response. Following an approach described in \cite{bm2008},  
we similarly binned two quantities useful for
imaging and spectroscopy: the effective exposure time and the estimate
of a background noise level. 
The effective exposure included the spatially variable mirror effective area and
detector quantum efficiency, the CCD gaps and bad pixels, and a
correction for the telescope motion. These quantities have been extracted
from the \chandra Calibration data base (CALDB 4.4.5) and the events
list. The background noise model consists of galactic foreground and
cosmic X-ray background (CXB) components, but also false detections
due to cosmic ray induced particles. It has been fit to the data
outside the region of the field of view covered by the target. As
described in \citealt{bartalucci} (in prep.), for each ACIS-I CCD 
the particle background model holds a spatially variable spectrum
including a power-law plus exponential continuum and several fluorescence
lines. It has been fit to \textit{very faint} out-of-focus and
blank-sky observations performed during the so-called D and E 
background periods, that cover our observation.

We have extracted X-ray images in the 0.5-2.5 keV energy band by correcting the
event images for effective exposure and background noise. The image
shown in Fig. \ref{fig:fig3} has been smoothed with a gaussian 
of $\sim$ 5\secpoint FWHM to reduce the shot noise, and covers 
the same region of the sky shown in Fig.\,\ref{fig:fig1} ({\em Left panel}). 
The \gmrt 323~MHz contours (same as in Fig. \ref{fig:fig_2}, {\em Left panel}) are overlaid on that. \\
Beyond the known X-ray emission associated with the galaxy cluster M\,0520  \citep{Ebeling10,Mantz10}, 
we clearly detect a SE diffuse X-ray  source, located about 8\minpoint from the centre of the main cluster and coincident 
with the newly detected diffuse radio source D1. The peak of X-ray emission of this extended 
structure corresponds to an unclassified source (namely 1WGA\,J0521.0-1333; hereafter 1WGA\,0521) 
in the WGA Catalog \citep{White94}. Compared to M\,0520, its surface brightness distribution 
appear to be much less regular, with the brightest part of the X-ray emission slightly elongated in the NE-SW direction.

\begin{figure*}
   \centering
   \includegraphics[width=0.75\textwidth]{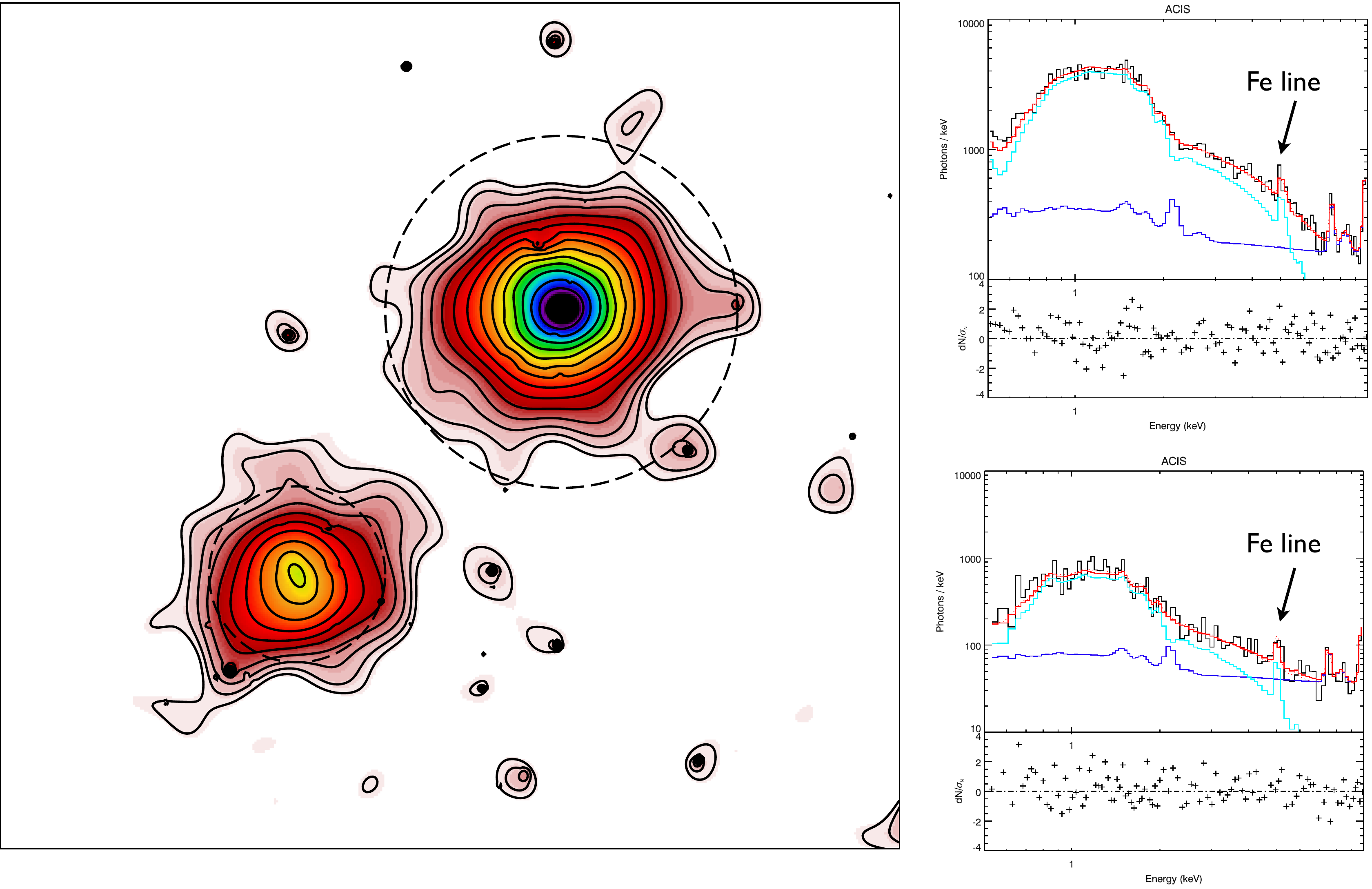}
      \caption{{\em Left panel:} Wavelet reconstructed \chandra X-ray image in the 0.5-2.5 keV energy band. The dashed circles correspond to 1 Mpc for M\,0520 and 0.5 Mpc for 1WGA\,0521, which have been used to extract the spectra.  {\em Right panel:} X-ray spectra of the two extended sources (top: M\,0520, bottom: 1WGA\,0521), extracted from the event list in the 0.5-10 keV band.} 
         \label{fig:fig4}
   \end{figure*}

The image shown in Fig. \ref{fig:fig4} ({\it Left panel}) 
is the result of a wavelet analysis including the denoising of the count rate map and the
restoration of the X-ray surface brightness. 
More precisely, the count
rate map has first been denoised from the 3 $\sigma$ thresholding of a
B3-spline undecimated wavelet transform. This was undertaken by
assuming the variance of the wavelet coefficients to be only dependent
of the effective exposure, thanks to the adaptation of a Multiscale
Variance Stabilization Transform, see details in \citealt{zhang08}.  
The exposure corrected background map has further been
projected into the same significant multi-resolution support \citep{murtagh95} 
and subtracted from the denoised count rate, yielding the X-ray surface brightness.

\subsection{Spectral analysis}
\label{sec:Xspec}

Spectral analyses have been performed assuming a redshifted and N$_H$
absorbed ICM emission spectrum modelled using the Astrophysical Plasma
Emission Code (APEC, \citealt{smith01}), with the element abundances of 
\cite{grevesse98} and neutral hydrogen absorption cross
sections of \cite{balucinska92}. The N$_H$ value has
been fixed to $7.3 \times 10^{24}$ m$^{-2}$, from measurements obtained
near M\,0520 in the Leiden /Argentine/Bonn Survey of galactic HI 
\citep{kalberla05}. ICM emission spectra are altered by the
effective exposure and background noise, but also convolved by a
redistribution function of the photon energies by the imaging
camera. First tabulated within 128 tails of the ACIS-I CCDs using the
\chandra Interactive Analysis of Observations (CIAO) software and CALDB
4.4.5, this function has been computed for each spectrum by 
averaging the redistribution functions associated with each event
position.

We have extracted the spectra of the two extended sources M\,0520 and
1WGA\,0521 from the event list in the 0.5--10 keV band, 
within two point-source-excluded circular regions centred on 
each of them, with radii of r=1 Mpc and r=0.5 Mpc, respectively. 
As shown on Fig. \ref{fig:fig4} (right panel), these spectra are ideally fit with two ICM emission
spectra including their iron line complex redshifted near 5 keV. 
This identifies both sources as galaxy clusters. 
Our spectral fit allowed us to estimate two redshift values of 0.336$^{+0.004}_{-0.002}$  and 0.34$^{+0.01}_{-0.02}$ 
as for M\,0520 and 1WGA\,0521, respectively, consistent with each other and with the 
redshift of M\,0520 measured from optical spectroscopy (see also Table \ref{tab:x}). 
Note that the radii of the two regions used to extract our spectra have been empirically chosen in order to maximise the signal
to noise ratio of the iron line, and thus best constrain these measurements. 
In the following estimates, we assume the redshift of both clusters to be fixed to this
optical redshift, z=0.336. To estimate the cluster masses within r$_{500}$ 
-- the radius of a sphere whose density is 500 times the critical
density of the universe -- , we computed their Y$_{X}$ parameter defined as
the product of gas mass M$_{g,500}$ and average temperature kT \citep{kravtsov06}.  
As described in \eg  \cite{bm2008}, we invert gas mass profiles
from the radially average surface brightness of each cluster, then
iterated about the Y$_{X}$-M$_{500}$ scaling relation calibrated from
hydrostatic mass estimates in a nearby sample of clusters observed
with \chandra (\citealt{vikhlinin09}). This yielded 
estimates of the total cluster masses M$_{500}$, radii r$_{500}$, gas
masses M$_{g,500}$, and average temperatures kT, that are reported in Table \ref{tab:x}.

%
\begin{table*}[ht!]
\caption{X-ray properties of the two galaxy clusters.}
\begin{center}
\begin{tabular}{lcccccccc}
\hline
Cluster & RA$_{J2000}$ & DEC$_{J2000}$ & z &  kT &  Y$_X$ &  M$_{g, 500}$ &  M$_{500}$ &  r$_{500}$ \\
	     & (h,m,s)  & (\degpoint, \minpoint, \secpoint)   & 	     & [keV]    &  [10$^{14}$M$_{\odot}$ keV]   &   [10$^{14}$ M$_{\odot}$]      & [10$^{14}$ M$_{\odot}$] & [kpc] \\
\hline
MACSJ\,0520--1328  	& 05 20 42.0 &  --13 28 50  & 0.336 $^{+0.004}_{-0.002}$ &  6.2 $^{+2.4}_{-1.2}$  & 4.0 $^{+1.6}_{-0.8}$   & 0.65 $^{+0.07}_{-0.05}$  & 5.3 $^{+1.5}_{- 0.8}$  & 1097 $^{+93}_{-60}$\\
1WGA\,J0521.0--1333 & 05 21 03.3  & --13 34 05  &  0.34 $^{+0.01}_{-0.02}$ &  3.6 $^{+1.4}_{-0.9}$  & 1.1 $^{+0.4}_{-0.3}$   & 0.29 $^{+0.04}_{-0.03}$  & 2.5 $^{+0.6}_{-0.4}$ & 851 $^{+63}_{-48}$\\
\hline
\end{tabular}
\end{center}
\label{tab:x}
\end{table*}

\section{Optical/IR analysis}
\label{sec:IRO}

In this Section we investigate the optical/IR properties of the galaxy 
clusters M0520 and 1WGA\,0521 
by using  available catalogs extracted from the WISE  and UKST Red public surveys  (\footnote{See \url{http://irsa.ipac.caltech.edu/Missions/wise.html} and \url{http://www-wfau.roe.ac.uk/sss/index.html}}). 

\subsection{Colour-magnitude diagrams}
\label{sec:cmd}

\begin{figure*}
   \centering
   \includegraphics[width=1.\textwidth]{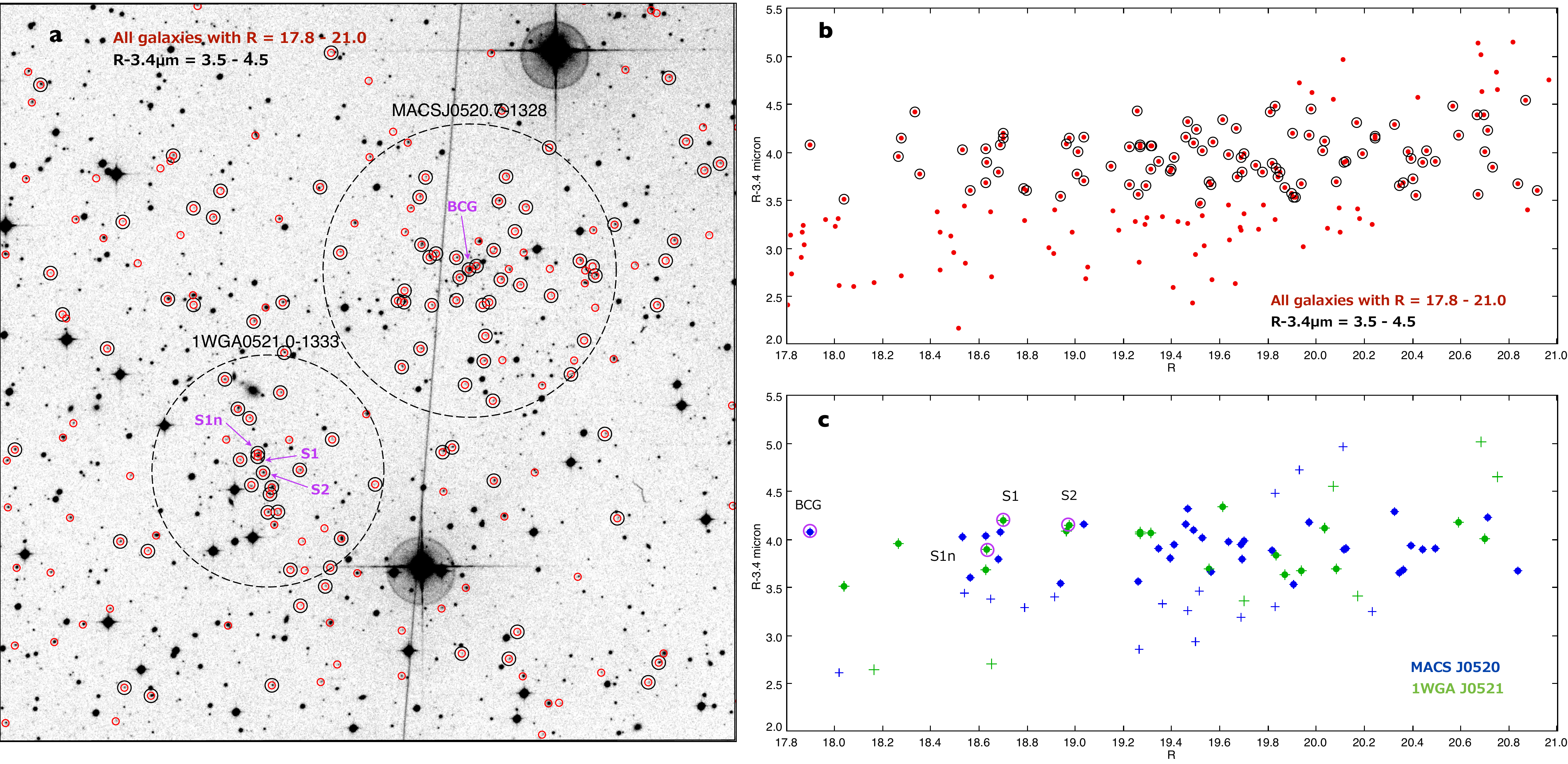}
      \caption{{\em panel a:}  DSS--R image of the  $\sim$19\minpoint$\times$19\minpoint area around M0520 (same as in Fig. \ref{fig:fig1}, {\em Right panel}); dashed circles have radii $r_{500}$, and are centred on the X-ray centre of each cluster. The red small circles mark the galaxies selected in the range R = 17.8 -- 21.0; the black large circles shows only those galaxies within the range of colours R -- 3.4 $\mu$m = 3.5 -- 4.5.  {\em panel b:} R -- 3.4 $\mu$m CMD of the galaxies within the  $\sim$19\minpoint$\times$19\minpoint area as in panel a: red and black circles as in panel a; 
{\em panel c:}  R -- 3.4 $\mu$m CMD of the galaxies  within the dashed circles in panel a: the crosses are all the galaxies in the range R = 17.8 -- 21.0, the filled dots are the colour-selected once (R -- 3.4 $\mu$m = 3.5 -- 4.5) .The BCG of M\,0520and the optical counterparts of radio sources S1 and S2 are marked by magenta circles (they are also highlighted by magenta arrows in panel a, for clarity). }
      \label{fig:fig5}
   \end{figure*}

Galaxy clusters are characterised by a well defined sequence of red galaxies, 
most likely early--type cluster members \citep[see \eg, ][]{gladders05}. \\ 
We have used multi-band optical and infrared data to trace the colour-magnitude diagram (hereafter CMD) 
of galaxies in the central field around M0520  (see panel a of Fig. \ref{fig:fig5}; $\sim$19\minpoint $\times$ 19\minpoint, same as in Fig. \ref{fig:fig1}, left). 
For this, we matched the archival WISE and UKST Red catalogs sources (with a tolerance radius of 6'' for association, \ie, $\sim$1 FWHM of the WISE band) 
and plotted the R--3.4 $\mu$m colour as a function of R magnitude. We selected only sources classified as galaxies based on their morphological parameters and areal profile shape \citep[``CLASS=1'' sources in SuperCOSMOS catalog,][]{Hambly01}. The CMD is shown in Fig. \ref{fig:fig5} (panels b and c). 
To trace it, we selected galaxies in the magnitude range R = 17.8 -- 21, \ie within approximately the BCG magnitude and the UKST SES-R survey magnitude limit \citep{Andernach99}. 
Despite  the fact that the field is dominated by cluster galaxies at a quite high-redshift for the magnitude limit of SuperCOSMOS catalog (R* $\sim$ 19.7 at $z \sim 0.336$), 
we identify the most-likely red-sequence region in the CMDs, in the range of colours $\sim$3.5-4.5 (see Fig. \ref{fig:fig5} (panel b), black circles). \\
Panel c of Fig. \ref{fig:fig5} shows the R -- 3.4 $\mu$m CMD for the galaxies located within circular areas of radii 1$r_{500}$ (see Table \ref{tab:x}) from the X-ray centre of each cluster (see dashed circles in Fig. \ref{fig:fig5}, a). As is clear from Fig. \ref{fig:fig5} (panel a),  almost all  the galaxies within these two cluster regions fall in the colour range $\sim$3.5-4.5, suggesting they all populate the most-likely identified red sequence region. \\
In particular, most of the objects located  in the 1WGA\,0521 region (green   in Fig. \ref{fig:fig5}, panel c) populate the same locus of the CMD diagrams as the BCG of M\,0520 and as  galaxies within the r$_{500}$ radius of the main cluster (blue  in Fig. \ref{fig:fig5}, panel c, and magenta circles).  
This is especially true for the likely optical counterparts of the two radio galaxies S1 and S2 (see Fig. \ref{fig:fig1}, {\em Right panel}). Two galaxies actually lies within the 10$\sigma$ contour around the radio peak of S, labelled as S1n and S1 in (see Fig. \ref{fig:fig5}, c); they both could be the optical counterpart of S1, and have similar colours.  \\
We have statistically compared the distribution in the CMD of the galaxies lying within the central regions of the two clusters (\ie the two circles shown in \ref{fig:fig5}, panel a). 
According to a 2D Kolmogorov--Smirnov test and with a significance level $\lesssim$5\%, we can exclude that the two galaxy distributions are drawn from a different population. 

\subsection{Galaxy isodensity map and multi--wavelength comparison}
\label{sec:iso}

\begin{figure*}
   \centering
   \includegraphics[width=0.6\textwidth]{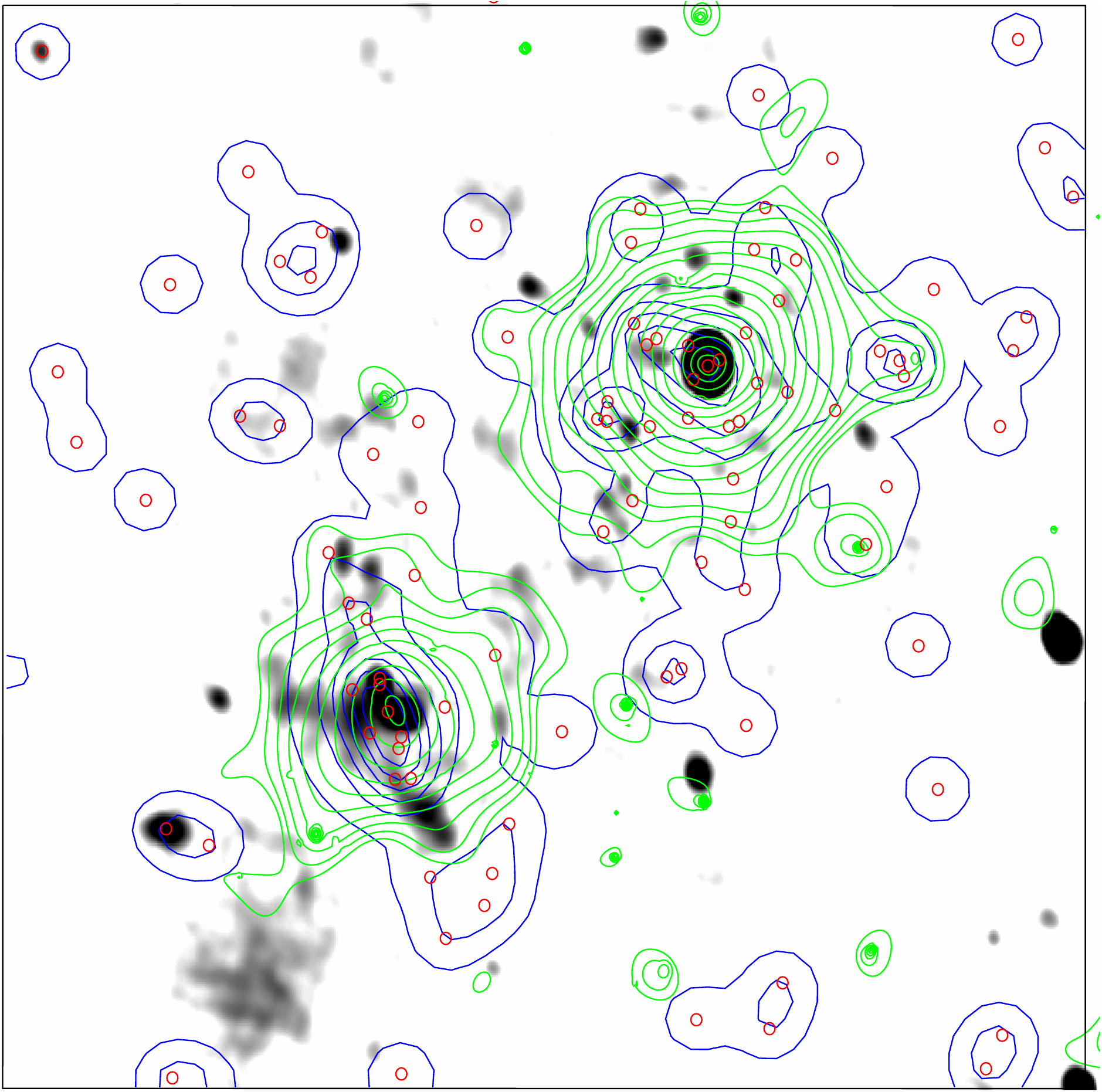}
      \caption{
      Galaxy isodensity map (blue contours), with overlaid the X-ray \chandra contours (green, same as in Fig. \ref{fig:fig4}). The radio image (same as in Fig. \ref{fig:fig_2}, left) is shown in greyscale (with the lowest level corresponding to 3$\sigma_{323\rm MHz}$=0.82 mJy/b). Only galaxies in the magnitude range R=(17.8--21) and in the colour range  3.4$\mu$m -- R = (3.5--4.5) have been used, and their spatial distribution is shown by red circles.
      }
      \label{fig:fig6}
   \end{figure*}

Figure \ref{fig:fig6} shows the galaxy iso--density map in the same 
19\minpoint $\times$ 19\minpoint centred in between the clusters M\,0520 and 1WGA\,0521. 
We have used the same catalogues  as for the CMD analysis, selecting 
only galaxies in the R magnitude range 17.8 -- 21.0 and in the 3.4$\mu$m--R colour range 3.5 -- 4.5 (from UKST and WISE; see Sec. \ref{sec:cmd}). 
The map has been derived on the basis of a multi-scale approach, as described in Ferrari et al. (2005),  
and has been optimised to point out large--scale substructures. 
The optical iso-density contours  (blue) are superposed on  
X-ray contours (green; same as in Fig. \ref{fig:fig4}, left), 
and  the 323 MHz \gmrt radio image (greyscale; from same image as in Fig. \ref{fig:fig_2} left). 
Red circles mark the position of all the galaxies in the catalogue used. \\
There is a very good match between the X-ray and optical emission of both clusters. 
M\,0520 appears to have a more regular, roundish shape (both in the X-ray and the optical), 
except in the central brightest X-ray region, where the iso--density contours  are elongated towards North--East.
On the other hand, the projected galaxy distribution of 1WGA\,0521 is far from being spherically symmetric, 
with a clear elongation in the South--West/North--East direction, resembling the X-ray surface brightness distribution (see Fig. \ref{fig:fig3}); 
in particular, most of the galaxies within its r$_{500}$ appears to be distributed along a very narrow line. 
Since it is expected the gas and  galaxies follow a similar distribution, this results support the idea 
that M\,0520 is a quite relaxed system, while the more irregular 1WGA\,0521 is a disturbed system. 

The diffuse radio source D1 is elongated in a similar direction as that of the main galaxy clump of 1WGA0521; 
the two diffuse components D2 and D3  partially match with the lowest optical iso-density contour of M\,0520. 
On the contrary, D4 is located in a outer region, covering an area where no X-ray emission 
is detected, nor significant concentration of galaxies are present (at the level of the survey's magnitude limit and of the \chandra exposure).

\section{Discussion and conclusions}

We have reported high sensitivity \gmrt\ radio observations at 323 MHz of the galaxy cluster M\,0520, 
that allowed the detection of complex diffuse radio emission. The main feature of this emission is located 
at a projected distance of $\sim$8' SE of the cluster centre and was found to coincide with an extended 
X-ray source seen in the archival \chandra image. 
We have performed a multi wavelength analysis based on archival X-ray and optical/IR data, 
that allowed us to classify it as a galaxy cluster, namely 1WGA\,0521. 

Our X-ray analysis allowed us to characterise this newly discovered cluster as a low--temperature, 
relatively small system.  
From the spectral fitting we derived a redshift that is consistent with the optical redshift of M\,0520, z=0.336. 
However the uncertainties on our X-ray estimate (Sec. \ref{sec:Xspec}, Table \ref{tab:x}) do not allow us to firmly conclude 
that the two clusters lie at the same distance
\footnote{
Note that, at z$\sim$0.336, the difference of 0.004 between the X-ray redshift estimates (see Table \ref{tab:x})
corresponds to 25 Mpc in luminosity distance. If we also consider their uncertainties from spectral fitting, the distance between the two clusters could be much larger.  
}. 
Despite the fact that at z=0.336  their projected separation of $\sim$2.3 Mpc would easily include the clusters virial radii, 
the  \chandra\ image does not show evidence of a bridge of X-ray emission between M\,0520 and 1WGA0521, 
that could be expected in case of interaction (though this may be also due to the low sensitivity of the exposure). \\
On the other hand, our optical/IR analysis supports the hypothesis that the two clusters lie at similar redshifts, 
since their galaxy populations lie in statistically similar regions of the colour magnitude diagram. 
Optical spectroscopic observations and deeper X-ray observations are needed to accurately determine the 1WGA\,0521 
redshift and to investigate the dynamical scenario of these two systems and possible ongoing physical interaction. \\
The optical iso-density map derived from candidate cluster members 
shows that 1WGA\,0521 has an elongated morphology, suggesting a disturbed dynamical state. 
Conversely, M\,0520 seems more relaxed (as found by \citealt{Ebeling10}), with a more regular shape, except for its inner region where 
the galaxy distribution is slightly elongated.

Our radio images at low resolution reveal a complex distribution of diffuse emission, that we have analysed  
by identifying four structures (namely D1, D2, D3 and D4, see Sect. \ref{sec:radio}). 
The diffuse elongated source D1 is the brightest of these features. It is characterised by large--scale 
 and  low--surface--brightness, and it is coincident with the newly detected, probably unrelaxed cluster of galaxy 1WGA\,0521.   
 These properties are typical for cluster--scale diffuse radio sources. 
Due to its projected location in the central region of the cluster, the emission could be a cluster radio halo 
with an elongated morphology. Such shape may also suggest the source is a radio relic 
(though located closer to the cluster centre with respect to what typically observed). 
The emission in this case would be associated with the passage of a shock wave, that may have either 
accelerated particles from the thermal ICM, or have revived fossil radio plasma from a previous activity of a radio--loud AGN 
(that could be one of the two nearby radio galaxies S1 and S2). 
Following the definition by \cite{kempner2004}, the source would be classified as \textit{radio gischt} in the former scenario, or a \textit{radio phoenix} in the latter. 
 The relic could be associated with the cluster M\,520, rather than to 1WGA\,0521. This hypothesis seems less likely, due to 
its large distance from the cluster centre ($\sim$ 2.3 Mpc), much larger than what is typically observed, and  
the fact that M\,520 seems to be a relaxed system, while relics are always found to be associated to merging clusters. 
Another possibility that cannot be firmly excluded is that D1 is actually the tail of a tailed radio galaxy, whose nucleus may be identified with S1 or S2.  
Our present radio data, however, do not allow us to disentangle between these possibilities; 
spectral and polarisation information is needed to clearly classify this source. \\ 
Beyond the prominent source D1, additional diffuse emission on large--scale is located around it (D2, D3, D4). 
The two elongated structures D2 and D3 are difficult to classify.  
 They have an even lower brightness, their shape is filamentary, 
and they are located in a region in between 1WGA0521 and M\,0520. 
D4 is even harder to interpret, due to its roundish shape and its location in a region where our present data 
does not show any significant optical or X-ray emission. 
We suggest that these features might be related to non-thermal components in the faint inter-galactic medium surrounding 1WGA0521, that 
is not detected by the current shallow X-ray data due to sensitivity limits, but may be revealed through more extensive observations (\eg \citealt{plck2013} and references therein). \\
Multi--frequency high sensitivity radio observations are essential to properly classify all the features of the complex extended radio emission,
 and to understand their origin. Follow up observations are planned. \\

Regardless the nature of the diffuse radio emission and its possible relation to the cluster's dynamical state, 
the most important result of our analysis is that our detection of large--scale diffuse radio emission 
has led us to the discovery of a new massive galaxy cluster, which was not previously known.  
Indeed, despite the fact that M0520 is a system in well studied cluster samples \citep{Ebeling10,Mantz10,Horesh10} and that our X-ray and optical/IR analyses are based on archival data, the existence of the neighbouring cluster 1WGA0521 has been ignored up to now. 
Although it is not uncommon that X-ray surveys miss clusters (see \eg \citealt{cagnoni01}), in this case the radio observations have been crucial to spot its existence. \\
Thanks to on-going and future deep all-sky radio surveys (e.g. LOFAR--Surveys, EMU--ASKAP, MeerKAT--MighTEE, MWA GLEAM), 
diffuse radio sources will become a powerful tool for the discovery of dynamically disturbed galaxy clusters and large--scale structure merging/accretion events.

\begin{acknowledgements}
We thank the anonymous referee for her/his useful comments. 
We are grateful to Gianfranco Brunetti and Monique Arnaud for helpful suggestions and fruitful discussions. 
We warmly thank Christophe Benoist for his help in deriving the galaxy isodensity map.  
We would like to thank the staff of the \gmrt that made these observations possible. 
\gmrt is run by the National Centre for Radio Astrophysics of the Tata Institute of Fundamental Research. 
The National Radio Astronomy Observatory is a facility of the National Science Foundation operated under cooperative agreement by Associated Universities, Inc. 
GM and CF acknowledge financial support by the ``{\it Agence Nationale de la Recherche}'' (ANR) through grant ANR-09-JCJC-0001-01.  GWP acknowledges  financial support by ANR  through grant ANR-11-BD56-015. 
\end{acknowledgements}

\bibliographystyle{aa}

\bibliography{mj0520_relic}

\end{document}